\documentclass[showpacs,preprintnumbers]{revtex4}
\usepackage{amssymb}
\usepackage{graphicx}
\usepackage{bm}
\usepackage{amsmath}
\usepackage{mathrsfs}

\begin{document}

\title{Linear-Optical Implementation of Perfect Discrimination between
Single-bit Unitary Operations}
\author{Pei Zhang}
\author{Liang Peng}
\author{Zhi-Wei Wang}
\author{Xi-Feng Ren}
\author{Bi-Heng Liu}
\author{Yun-Feng Huang \footnote[1]{Corresponding author: hyf@ustc.edu.cn}}
\author{Guang-Can Guo}
\address{Key Laboratory of Quantum Information, University of Science and
Technology of China, CAS, Hefei 230026, People's Republic of China}

\begin{abstract}
Discrimination of unitary operations is a fundamental quantum
information processing task. Assisted with linear optical elements,
we experimentally demonstrate perfect discrimination between
single-bit unitary operations using the sequential scheme which is
proved by Runyao Duan \textit{et al.} [Phys. Rev. Lett. 98, 100503
(2007)]. We also make a comparison with another perfect
discrimination scheme named parallel scheme. The complexity and
resource consumed are analyzed.
\end{abstract}

 \maketitle


For quantum computing and quantum information processing, one important task
is the discrimination of quantum states and unitary operations. This is
strongly related to quantum nonorthogonality, which is one of the basic
features of quantum mechanics. Since the pioneering work of Helstrom \cite%
{Helstrom} on quantum hypothesis testing, the problem of
discriminating nonorthogonal quantum states has attracted much
attention, both in theory \cite{Chefles, Bergou} and experiments
\cite{Huttner, Barnett, Clarke, Kendon, Mohseni}. Naturally, the
concepts of nonorthogonality and distinguishability can also be
applied to quantum operations. However, things become very different
when we refer to perfect discrimination of quantum operations which
already has some theoretical works \cite{Acin, D, Duan, Zhou,
Vertesi, Duan1}. Some pioneering works have been devoted to a good
understanding of the exact role of quantum entanglement in the
discrimination between unitary operations. It has been proved
\cite{Acin, D} that perfect discrimination of nonorthogonal unitary
operations can always be achieved with a finite number of running
the unknown unitary operation, by using a suitable entangled state
as input. This is contrary to the case of nonorthogonal quantum
states, for which perfect discrimination can not be achieved with
finite number of copies. Recently, Runyao Duan et al indicated that
entanglement is not necessary for discrimination between unitary
operations \cite{Duan}. They show that by taking a suitable state
and proper auxiliary unitary operation, nonorthogonal unitary
operations can also be perfectly discriminated even without
entanglement. This result impacts on the role of quantum
entanglement in the context of quantum computing \cite{Meyer}, and
also makes experiment much easier because no $N$-partite entangled
state is needed.

First, we briefly explain the basic idea of perfect discrimination
between unitary operations \cite{Acin, D, Duan}. Suppose we have an
unknown quantum circuit which is secretly chosen from two
alternatives, $U$ and $V$. Here both $U$ and $V$ are unitary
operations acting on a $d$-dimensional Hilbert space
$\mathscr{H}_{d}$. We apply the unknown circuit to a proper initial
state $\left\vert \psi _{i}\right\rangle $. If the output states
$\left\vert \psi _{o}\right\rangle _{U}$ and $\left\vert \psi
_{o}\right\rangle _{V}$ are orthogonal, then perfect discrimination
between $U$ and $V$ is achieved. By denoting the length of the
smallest arc containing all the eigenvalues of $U$ on the unit
circle by $\Theta (U)$, $U$ and $V$ can be perfectly distinguished
if and only if $\Theta (U^{\dag }V)\geqslant \pi $. The minimal $N$
such that $\Theta ((U^{\dag }V)^{\otimes N})\geqslant \pi $ is given
by $\lceil \frac{\pi }{\Theta (U^{\dag }V)}\rceil $. Here $\lceil
x\rceil $ denotes the smallest integer that is not less than $x.$ To
determine what the circuit really is, we can use the parallel scheme
which is brought forward in Ref. \cite{Acin, D} or sequential scheme
which is brought forward in Ref. \cite{Duan}.

\begin{figure}[tbh]
\includegraphics[width=9cm]{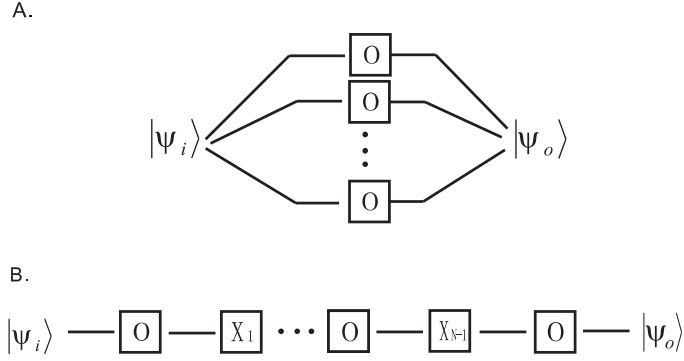}
\caption{Two schemes of perfect discrimination two unitary operations. $O$
represents one of the unknown circuit, $U$ or $V$. $\left\vert \protect\psi %
_{i}\right\rangle $ is the input state and $\left\vert \protect\psi %
_{o}\right\rangle $ is the output state. Figure A: A parallel scheme for
discriminating two unitary operations U and V. Figure B: A sequential scheme
for discriminating two unitary operations U and V. $X_{1},X_{2}...,X_{N-1}$
are auxiliary unitary operations. $\left\vert \protect\psi _{i}\right\rangle
$ is the input state and $\left\vert \protect\psi _{o}\right\rangle $ is the
output state. The output states $\left\vert \protect\psi _{o}\right\rangle
_{U}$ and $\left\vert \protect\psi _{o}\right\rangle _{V}$\ need to be
orthogonal for perfect discrimination.}
\end{figure}

For the parallel scheme, shown in Fig. 1A, we can perfectly discriminate $U$
and $V$ by three steps: First, prepare a proper $N$-qudit entangled state $%
\left\vert \psi _{i}\right\rangle $ ; second, apply the secretly chosen
circuit $N$ times on $\left\vert \psi _{i}\right\rangle $, where each qudit
one time; last, perform a projective measurement on the output states $%
\left\vert \psi _{o}\right\rangle _{U}=U^{\otimes N}\left\vert \psi
_{i}\right\rangle $ and $\left\vert \psi _{o}\right\rangle
_{V}=V^{\otimes N}\left\vert \psi _{i}\right\rangle $. If
$\left\vert \psi _{o}\right\rangle _{U}\perp \left\vert \psi
_{o}\right\rangle _{V}$ (``$\perp$" represents the orthogonality of
two states), then two unitary operations are perfect discriminated.
Indeed, we can always find a proper input state to make sure
$\left\vert \psi _{o}\right\rangle _{U}\perp \left\vert \psi
_{o}\right\rangle _{V}$ by using this method \cite{Acin, D}.

While for the sequential scheme, shown in Fig. 1B, we need to prepare a
proper one qudit state $\left\vert \psi _{i}\right\rangle \in \mathscr{H}%
_{d} $ and perform the unknown circuit sequentially on $\left\vert \psi
_{i}\right\rangle $ for $N$ times. Between each two runs of the unknown
circuits, a proper auxiliary unitary operation $X_{i}\in \mathscr{U}$ ($%
i=1,...,N-1$) is inserted. Here $\mathscr{U}$ represents a set of
unitary operations acting on $\mathscr{H}_{d}$. Perfect
discrimination between $U$ and $V$ requires that two output states
are orthogonal, such that
\begin{equation}
UX_{N-1}\cdot \cdot \cdot UX_{1}U\left\vert \psi _{i}\right\rangle
\perp VX_{N-1}\cdot \cdot \cdot VX_{1}V\left\vert \psi
_{i}\right\rangle .
\end{equation}%
As proved in Ref. \cite{Duan}, this discrimination task can always
be translated to distinguish $U^{\dag }V$ and identity ($I$). Then,
relation (1) can be reduced as follows:
\begin{equation}
(U^{\dag }V)X(U^{\dag }V)^{N-1}\left\vert \psi _{i}\right\rangle \perp
X\left\vert \psi _{i}\right\rangle .
\end{equation}%
Where $X_{i}$, $X$ and $\left\vert \psi _{i}\right\rangle $ depend on $%
\Theta (U^{\dag }V)$.

By comparison of these two schemes, sequential scheme is
experimentally much easier than parallel scheme especially in
testing the orthogonality of output states. For sequential scheme,
only single qubit projective measurement is needed while for
parallel scheme non-local measurements should be performed, which is
rather difficult within current technology. Further more, because of
the environment noise and experimental error, the output states of
parallel scheme are always mixed entangled states which are hardly
precisely distinguished by local operation and classical
communication (LOCC) \cite{Peres}.

In this paper, we report an experiment which demonstrates perfect
discrimination between single-qubit unitary operations using the
sequential scheme \cite{Duan}. Linear optical elements are used to
perform unitary operations on photonic qubits in the experiment. At
the end we perform a projective measurement to show the perfect
discrimination. The quality of our unitary operation is
characterized by the average precess fidelity $F(U)$
\cite{Altepeter, Wang} between the matrix which is got from standard
quantum process tomography \cite{Poyatos, Chuang, Nielsen, Childs,
Mitchell} and the theoretical one.

Although the sequential scheme is not limited to single-qubit unitary
operations, we choose to discriminate two single-bit unitary operations in
our experiment. Because single-qubit unitary operation is a class of
fundamental unitary transformations which is widely used in quantum
computing and quantum information processing tasks. And with linear optical
elements, any single-qubit unitary operation on photonic polarization qubit
can be easily implemented using a set of wave-plates \cite{Englert}. For
simplicity, we set the run times $N$ of the unknown circuit to be $N=\lceil
\frac{\pi }{\Theta (U^{+}V)}\rceil =2.$ And two sets of unitary operations
are chose:
\begin{equation}
Case1: U_{1}=\left(
\begin{array}{cc}
e^{i\frac{2}{3}\pi } & 0 \\
0 & 1%
\end{array}%
\right) , V_{1}=\left(
\begin{array}{cc}
e^{i\frac{1}{6}\pi } & 0 \\
0 & 1%
\end{array}%
\right) , \Theta (U_{1}^{\dag }V_{1})=\frac{\pi }{2}
\end{equation}
\begin{equation}
Case2: U_{2}=U_{1}=\left(
\begin{array}{cc}
e^{i\frac{2}{3}\pi } & 0 \\
0 & 1%
\end{array}%
\right) , V_{2}=I=\left(
\begin{array}{cc}
1 & 0 \\
0 & 1%
\end{array}%
\right) , \Theta (U_{2}^{\dag }V_{2})=\frac{2\pi }{3}.
\end{equation}

Case 1: $N=\lceil \frac{\pi }{\Theta (U_{1}^{\dag }V_{1})}\rceil =2$,
auxiliary operation $X=\left(
\begin{array}{cc}
\cos \alpha & -\sin \alpha \\
\sin \alpha & \cos \alpha%
\end{array}%
\right) =I$, where $\alpha =\arctan \sqrt{-\frac{\cos (N\Theta /2)}{\cos
((N-2)\Theta /2)}}$, and $\left\vert \psi _{i}\right\rangle =\frac{1}{\sqrt{2%
}}(\varphi _{_{1}}+\varphi _{_{2}})=\frac{1}{\sqrt{2}}(\left\vert
H\right\rangle +\left\vert V\right\rangle ),$ where $\varphi _{_{1}}$ and $%
\varphi _{_{2}}$ are the eigenvectors of $X^{\dag }(U^{\dag
}V)X(U^{\dag }V)^{N-1}$ \cite{Duan}. The experimental setup is shown
in Fig. 2. We do a projective measurement to show the results of
perfect discrimination. The
projective basis we chose is $\left\vert \psi _{b}\right\rangle =\frac{1}{%
\sqrt{2}}(-e^{\frac{\pi }{3}i}\left\vert H\right\rangle +\left\vert
V\right\rangle )$. Photons will be only detected by D$_{1}$ if their
state is $\left\vert \psi _{b}\right\rangle$ or will be only detect
by D$_{2}$ if their state is the orthogonal state of $\left\vert
\psi _{b}\right\rangle$. All the operations are achieved by linear
optical components. And all the degrees of wave plates are shown in
Table I. We do a process tomography of our unitary operations to
ensure their
validity. The average process fidelities we got are $F(U_{1})=0.985$ and $%
F(V_{1})=0.975$. After a theoretical calculation, we know that
the output state $\left\vert \psi _{o}\right\rangle$ is $\frac{1}{%
\sqrt{2}}(-e^{\frac{\pi }{3}i}\left\vert H\right\rangle +\left\vert
V\right\rangle )$ when applying $U_{1}$ while
$\left\vert \psi' _{o}\right\rangle$ is $\frac{1}{%
\sqrt{2}}(-e^{\frac{\pi }{3}i}\left\vert H\right\rangle -\left\vert
V\right\rangle )$ when applying $V_{1}$, and $\langle \psi _{o}|
 \psi' _{o}\rangle=0$. The polarized beam
splitter (PBS) transmits vertical polarizing photons and reflects
horizontal polarizing photons. So, with our projective measurement,
we get the results that the unknown unitary operation is $U_{1}$
when photons are only detected by D$_{1}$ and the unknown unitary
operation is $V_{1}$ when photons are only detected by D$_{2}$. The
results are shown in Fig. 3.

Case 2: Similar to the discussions in Case 1, we get $N=\lceil
\frac{\pi }{\Theta
(U_{2}^{\dag }V_{2})}\rceil =2$, auxiliary operation $X=\frac{1}{\sqrt{3}}%
\left(
\begin{array}{cc}
\sqrt{2} & -1 \\
1 & \sqrt{2}%
\end{array}%
\right) $ and $\left\vert \psi _{i}\right\rangle =
(-0.151+0.262i)\left\vert H\right\rangle +0.953\left\vert
V\right\rangle )$. The
projective basis we chose is $\left\vert \psi _{b}\right\rangle =\frac{1}{%
\sqrt{2}}(e^{-\frac{\pi }{6}i}\left\vert H\right\rangle +\left\vert
V\right\rangle )$. The experimental setup is not changed but the
settings of wave plates are different from the settings in Case 1.
All the settings of wave plates are shown in Table I. The average
process fidelity of the auxiliary operation $X$ is $F(X)=0.996$. We
can get two orthogonal output states $\left\vert \psi
_{o}\right\rangle=\frac{1}{%
\sqrt{2}}(e^{-\frac{\pi }{6}i}\left\vert H\right\rangle +\left\vert
V\right\rangle )$ and $\left\vert \psi'
_{o}\right\rangle=(\frac{1}{%
\sqrt{2}}(e^{-\frac{\pi }{6}i}\left\vert H\right\rangle -\left\vert
V\right\rangle )$ when applying $U_{2}$ and $V_{2}$ respectively. So
we know that the unknown unitary operation is $U_{2}$ when photons
are only detected by D$_{1}$ and the unknown unitary operation is
$V_{2}$ when photons are only detected by D$_{2}$ which are shown in
Fig. 3.

\begin{figure}[tbh]
\includegraphics[width=14cm]{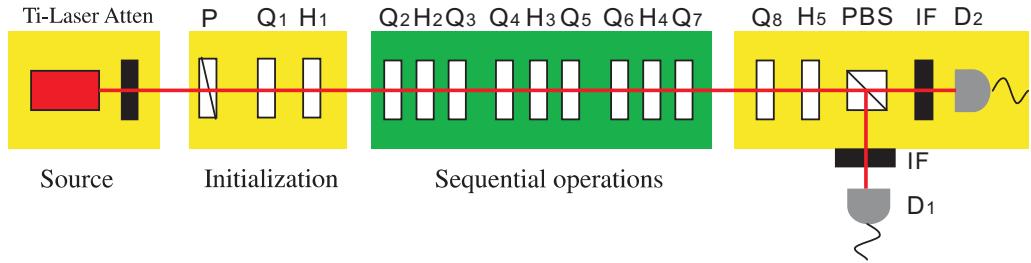}
\caption{Experimental setup for sequential scheme. Atten represents
the attenuator. P is the polarizer. PBS represents the polarized
beam splitter which transmits vertical polarizing photons and
reflects horizontal polarizing photons. IF is interference filter
centred at $702nm$ with $4nm$ bandwidth. D$_{1}$ and D$_{2}$ are
photon detectors. Q$_{i}$ ($i=1,2...8$) and H$_{j}$ ($j=1,2...5$)
represent quarter wave plates (QWP) and half wave plates (HWP)
respectively. Our unitary operations ($U$, $V$ and $X$) are all
realized by a HWP and two QWPs. At the end, we perform the
projective measurement.}
\end{figure}

\begin{table}[h]
\caption{Wave plates setting (degree)}\tabcolsep 3.5mm
\par
\begin{center}
\begin{tabular}{|r||r|r||r|r|r||r|r|r||r|r|r||r|r|}
\hline & \multicolumn{2}{|c||}{input} & \multicolumn{3}{|c||}{U(V)}
& \multicolumn{3}{|c||}{X} & \multicolumn{3}{|c||}{U(V)} &
\multicolumn{2}{|c|}{measurement} \\ \cline{2-14} & Q$_{1}$ &
H$_{1}$ & Q$_{2}$ & H$_{2}$ & Q$_{3}$ & Q$_{4}$ & H$_{3}$ & Q$_{5}$ & Q$_{6}$ & H$_{4}$ & Q$_{7}$ & Q$_{8}$ & H$_{5}$ \\
\hline
U$_{1}$ & 0 & 22.5 & 45 & 15 & 45 & 0 & 0 & 0 & 45 & 15 & 45 & 45 & 37.5 \\
\hline V$_{1}$ & 0 & 22.5 & 45 & 37.5 & 45 & 0 & 0 & 0 & 45 & 37.5 &
45 & 45 & 37.5
\\ \hline
U$_{2}$ & -15 & 42.4 & 45 & 15 & 45 & 27.4 & 45 & 62.6 & 45 & 15 & 45 & 45 & 15 \\
\hline V$_{2}$ & -15 & 42.4 & 0 & 0 & 0 & 27.4 & 45 & 62.6 & 0 & 0 & 0 & 45 & 15 \\
\hline
\end{tabular}%
\end{center}
\end{table}

In our experimental setup, the single photon source is achieved by
by attenuating the coherent Ti-sapphire laser to single photon
level. Then a polarizer (transmitted horizontal polarizing photon),
quarter wave plate (QWP) Q$_{1}$ and half wave plate (HWP) H$_{1}$
make up the initialization of input states. The set of wave plates
with one half wave plate sandwiched by two quarter wave plates can
realize any unitary operation on the photonic polarization qubit,
which can be represented in terms of three Eulerian angles.
Projective measurement consists of QWP (Q$_{8}$), HWP (H$_{5}$), PBS
(transmitting vertical polarizing photons) and two photon detectors.
It is used to verify the orthogonality of output states.

\begin{figure}[tbh]
\includegraphics[width=11cm]{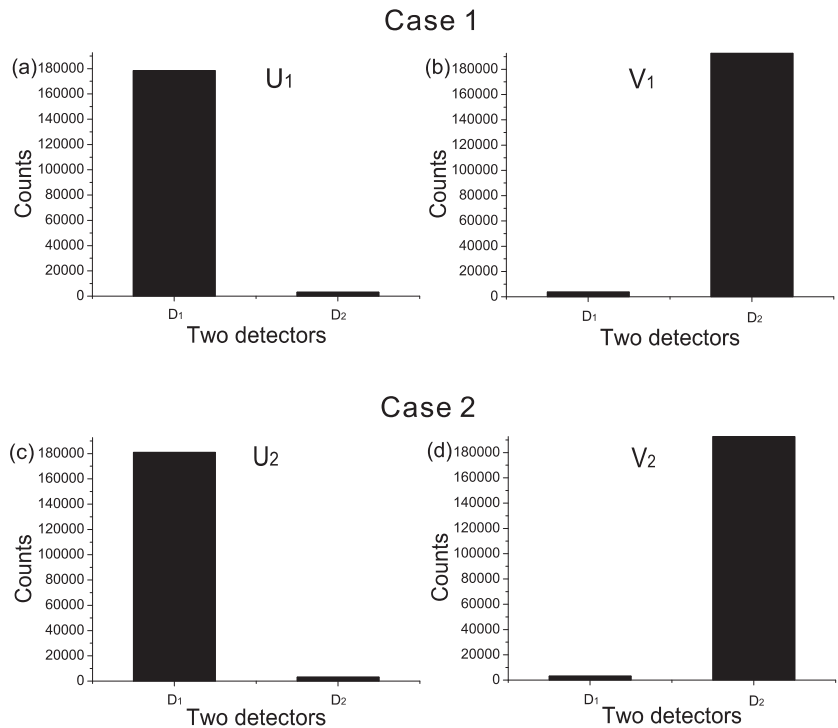}
\caption{Experimental results of perfect discrimination. The above
two figures are the results of Case 1, and the bottom two figures
are the results of Case 2. We can directly discriminate the circuits
by counts of two detectors.}
\end{figure}

Our results are shown in Fig. 3. When choosing the input states and
projective measurements discussed above, we can discriminate the
circuits via the results of photon detectors. The probabilities of
successful discrimination are about $98.0\%$ \cite{25} (Fig 3(a)),
$98.1\%$ (Fig 3(b)) of Case 1, and $98.3\%$ (Fig 3(c)), $98.4\%$
(Fig 3(d)) of Case 2. In our results, the errors mainly come from
the deviations of the angle settings of wave plates. Because there
are many wave plates in our experimental setup and the precision of
them is only about $0.2^\circ$. This leads to the discrimination
probability less than $1$. But it is quite different from other
non-perfect discrimination schemes whose
probabilities of successful discrimination never achieve $%
100\% $, even in theory.

We have experimentally distinguished two sets of single-bit unitary
operations where $N=2$ using sequential scheme. If $N>2$, we need to
apply unknown unitary operation $N$ times and auxiliary unitary
operations $N-1$ times on the input state. This is easy to be
implemented compared with the parallel scheme mentioned above.
However, it needs to perform sequentially at least $2N-1$ steps of
unitary operations. This may lead to a long discriminating time,
meanwhile, the deviations of the results would be influenced by the
additional $N-1$ auxiliary unitary operations. Instead, in the
parallel scheme one needs to prepare an $N$-partite entangled state
as the input state. When we have at least $N$ copies of the unknown
circuit and a suitable entangled resource, we can complete the
discrimination within a single step by applying $N$ copies of the
unknown circuit to the input state. This scheme is fast (only one
step) but costs entanglement resource, and there are also some
practical difficulties in non-local measurements.

\acknowledgments The authors thank Runyao Duan and Xiang-Fa Zhou for
interesting and helpful discussions. This work was funded by the National
Fundamental Research Program, National Natural Science Foundation of China
(Grant No. 60121503 and Grant No. 10674128), Innovation Funds from Chinese
Academy of Sciences, Program for New Century Excellent Talents in
University, A Foundation for the Author of National Excellent Doctoral
Dissertation of PR China (Grant No. 200729), and Doctor Foundation of
Education Ministry of China.


\begin{thebibliography}{99}
\bibitem{25} Supose the counts of detector D1 and D2 are $c1$ and $c2$,
respectively. Then the successful discrimination probability can be
written as $p=c1/(c1+c2)$ or $p=c2/(c1+c2)$ based on projective
basis.
\end{thebibliography}
\end{document}